# A Constructive Heuristic Algorithm for 3D Bin Packing of Irregular Shaped Items

Qiruyi Zuo[1], Xinglu Liu[1], and Wai Kin Victor Chan [*1]

[1] Intelligent Transportation and Logistics Systems Laboratory, Tsinghua-Berkeley Shenzhen Institute, Tsinghua University, Shenzhen 518055, China

**Abstract.** The three-dimensional bin packing problem (3D-BPP) plays an important role in city logistics and manufacturing environments, due to its direct relevance to operational cost. Most existing literature have investigated the conventional 3D-BPP, in which the shape of items are typically considered as regular shapes, e.g., rectangular-shaped rigid boxes or cylindrical-shaped containers. However, 3D-BPP for non-rectangular shaped items are quite common in varies delivery schemes, especially in fresh food delivery, and few published studies focusing on these issues. In this paper, we address a novel 3D-BPP variant in which the shape changing factor of non-rectangular and deformable items is incorporated to further enhance the loading efficiency and reduce the operational cost of related companies. Motivated by the compression process of item-loading, we propose a constructive heuristic (i.e., an improved dynamic-volume-based packing algorithm) to solve the studied problem. Experimental results over a set of randomly generated instances reveal that considering shape changing factor is indeed able to achieve higher space utilization than that of conventional schemes, thereby has potential to save packaging and delivering cost, as well as enhance operation efficiency.

## 1. INTRODUCTION

Three-dimensional bin packing problem(3D-BPP) is critical for those supply chain and logistics companies with massive delivery services due to its direct relevance with operational cost [1], e.g., JD.com, CAINIAO, SF Express, Amazon, etc. According to a news released by a Chinese media[2], CAINIAO applies a packing algorithm to create packaging materials saving by 15%, achieving a reduction of 290 million in express parcel amount. This successful application demonstrates that efficient packing solutions have huge potential to achieve considerable operational cost savings for companies in real operations. Moreover, 3D-BPP attracts numerous research efforts from scholars, e.g., Martello et al. [2], Lodi et al. [3], Gzara et al. [4] and Jiang et al. [5].

Although the published literature focusing on optimizing packing solutions is extensive, most of which consider the items are regular-shaped (e.g., rectangular-shaped or cylindrical-shaped). However, irregular-shaped item packing is ubiquitous in delivery and storage environments, especially for fresh food picking and packing (e.g., Amazon fresh[3], Food For Free[4], fruitrunner[5] and grocery industry (e.g., Misfits Market[6], Goteso[7], Delish[8]). Fig. 1.(a-c) provides an actual example of packing irregular-shaped items. These full packed boxes will be delivered to the customers (end customer or downstream retailers) after packing procedure.

Solving the conventional 3D-BPP is already extremely challenging due to its NP-hardness nature. Therefore, the problem is even harder to handle after incorporating the characteristic of irregular shape. Currently, the aforementioned irregular-shaped item packing tasks in real schemes are typically performed by workers without providing any reference packing solutions for them in advance (as Fig. 1.(b) shows), i.e., these packing operations have not yet well optimized. This inevitably leads to huge container space waste, low packing efficiency, as well as high packaging cost and labor cost.

---

[*] Corresponding author: Wai Kin Victor Chan, E-mail: chanw@sz.tsinghua.edu.cn
[2] VIPUTRANS, https://www.viputrans.com/cainiao-network-exploring-new-models-of-green
[3] https://www.amazon.com/
[4] https://foodforfree.org/
[5] https://www.fruitrunner.co.uk/
[6] https://www.misfitsmarket.com/
[7] https://www.goteso.com/
[8] https://www.delish.com/

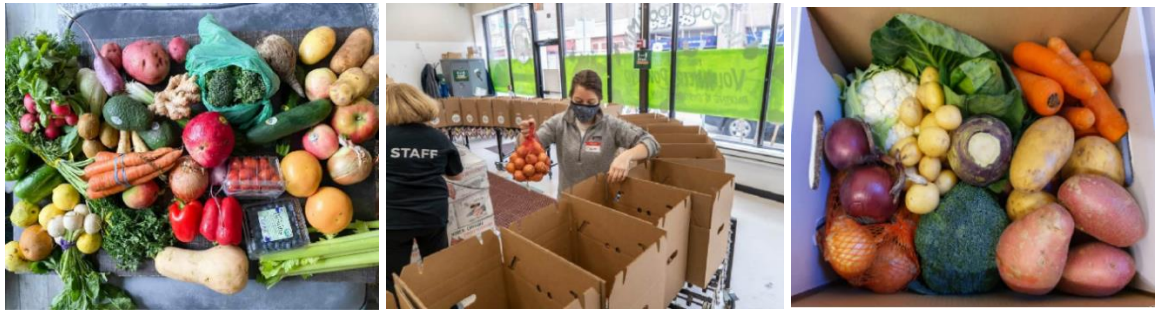

(a) To-be-packed irregular shaped items[1]   (b) Packing environments[2]   (c) Packing results[3]

**Fig. 1.** Packing irregular shaped items into bins.

To cope with these issues, in this work, we first consider the dynamic volume change of irregular-shaped items in 3D-BPP, and propose a constructive heuristic algorithm (i.e., dynamic-volume-based packing algorithm) to solve it. We then conduct a series of preliminary numerical experiments to investigate the benefits of volume changing factor. Furthermore, we offer workers with visualized packing sequence to help them accomplish the optimized packing solutions more easily. Results on the tested instances indicates that considering shape changing issue can create significant space utilization improvement compared with the problem under conventional constant volume and shape assumptions. Consequently, developing more realistic approaches for irregular item packing is propose to further save packaging and delivering cost, as well as enhance operation efficiency.

## 2. RELATED WORK

The 3D-BPP is NP-hard and thus is extremely hard to solve. Existing literature on this problem mainly focus on various problem variants and solution approaches. From the perspective of problem characteristics, research issues on this topic can be categorized into two groups: i) regular-shaped item packing; ii) irregular-shaped item packing.

Regular-shape-focused literature typically aim to load rectangular and cylinder items into bins. They either design exact [6, 7, 8, 9, 10, 11] or heuristic [12, 13, 14, 15, 16, 17] algorithms to yield high quality solutions. In addition, several papers attempt to apply reinforcement learning techniques to handle 3D BPP, e.g., Jiang et al. [18].

Although the previously mentioned studies are able to enhance the loading efficiency, some realistic features are failed to be contained yet, for example, irregular factor, rotations. Irregular factor refers to item shape and volume. Several published studies have addressed the packing under irregular shape consideration, e.g., Martinez-Sykora et al. [19], Abeysooriya et al. [20]. However, each item is assumed to be rigid and its volume and shape are constant over the entire packing process. This assumption is reasonable for well-packaged express parcels or those hard objects, but it is too ideal for soft items, e.g., fresh food. In fact, the item's actual volume usually changes dynamically as the loading process proceeds, e,g., packing vegetables or groceries into bins. More specifically, the occupied space of an item will reduce if there is pressure from other objects. Ignoring volume changing factor will cause possible space and packaging materials waste to some extent. Therefore, making volume change assumption is propose to create benefits on cost reduction. Nevertheless, to the best of our knowledge, no existing research address this problem. This work maybe the first attempt to consider dynamic volume change in the context of 3D-BPP and develop corresponding solving algorithm.

## 3. PROBLEM DISCRIPTION AND FORMULATION

### 3.1. PRELIMINARIES

---

[1] Misfits Market Review and Latest Coupons - Trial and Eater
[2] Going the extra miles: Food For Free's Just Eats grocery boxes head to the North Shore - FoodForFree.org
[3] Fruit and Veg Box - fruitrunner

Before detailing the mathematical descriptions, we first begin with a real scheme. Consider a fresh food wholesale delivery company aims to pack a batch of different kinds of fresh food into a number of rectangular bins and then deliver these bins to the customers. Fig. 2. shows a particle order received from some customer (e.g., restaurants, supermarkets). The company receives numerous orders from various customers every day. To accomplish these orders, the company first purchases foods and then arrange works to picking and packing foods into bins. Then a fleet of vehicles are dispatched to deliver the packed bins to the customers.

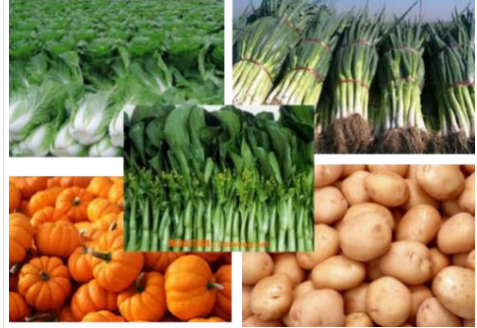

**Fig. 2.** A customer order of fresh wholesalers containing some vegetables and fruits.

To improve the delivery efficiency, we make the following assumptions according to practical requirements:

1) Fresh food for different customers cannot be packed in the same bin.

2) The same kind of fresh food will be placed in the same bin first.

In addition, to save container (basket) rental cost, companies usually aim to minimize the number of needed bins. In this work, we focus on a simple reduced version of this problem, i.e., we only attempt to maximize the initial volume of all the items of each single bin, which also typically results in greater space utilization.

### 3.2. MATHEMATICAL DESCRIPTIONS

At present, numerous published papers investigate the bin packing problem of two-dimensional irregular shapes. The main methods of solving these problems are approximation and heuristic, e.g., envelope method [21], graph theory method [22], etc. To be more specific, the main idea of rectangle envelope method is to approximate the combination of single or multiple parts with the minimum envelope rectangle, and then convert it into a rectangular-shaped item packing problem. Such approach is simple and easy to implement for real cases.

Motivated by the rectangular envelope method in two-dimensional bin packing problem, we approximate three-dimensional irregular-shaped item as the smallest envelope cube, as shown in Fig. 3. This approximation method is simple and easy to implement, but usually suffers from large errors and cannot accurately describe the size of the item. With this in mind, we introduce the process of item compression. Through this method, we take into account the possible compression characteristics of some special objects in real life (such as most of the vegetables and fruits in fresh food), and reduce the volume error after approximation in the meantime. In our work, we consider the irregular-shaped items are deformable and their height will be compressed during the packing process. Therefore, the actual occupied volume of packed items changes dynamically while executing the packing process. But the volume change must lie in reasonable range. If not, the items (foods, vegetables) will be crushed and cannot be used anymore. To avoid this issue, we need to restrict that the actual compressed height ratio is smaller than a maximum compression ratio. Based on the above interpretations, we can first approximate the to-be-packed items with rectangles, and then transfer the 3D irregular-shaped item packing problem into a 3D regular-shaped item packing problem with volume change consideration.

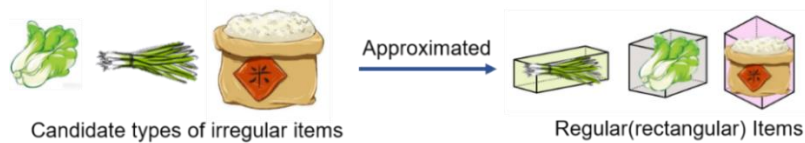

**Fig. 3.** The irregular item is approximated by a regular item which is the corresponding smallest envelope cube.

Now let us present a formal mathematical description of 3D-BPP for irregular-shaped items. Consider there are a given set of $n$ three-dimensional, rectangular and deformable items, $I = \{I_1, I_2, ..., I_n\}$. Each item in the set $I$ is characterized by its depth $d_i$, width $w_i$, initial height $h_i$, weight $m_i$, compressibility $c_i$ and maximum compression ratio $r_i$, $\forall i = \{1, 2, ..., n\}$. Compressibility indicates the proportion of deformation of an item due to the total weight of the items above it. The maximum compression ratio indicates the maximum ratio value by which the height of the item can be reduced. This parameter is to avoid over-squeezing and ensure the item can still be utilized. During the process of bin packing, there are also some properties of current item that need to be recorded, including true height after being packed $h_i^*$, true compression ratio $c_i^*$ and the total weight $u_i$ of all items placed on top of current item. The relations among the characteristics and properties mentioned above satisfy following equations:

$$h_i^* = h_i(1 - c_i^*), \quad \forall i = \{1, 2, ..., n\} \tag{1}$$

$$c_i^* = \min\left\{r_i, \frac{c_i u_i}{m_i}\right\}, \quad \forall i = \{1, 2, ..., n\} \tag{2}$$

The items are packed into a number of rectangular bins in $k$ different sizes with dimensions of depth $D_j$, width $W_j$ and height $H_j$. The upper face of each bin is opened and it is located in a three-dimensional coordinate system with the back-left-bottom corner in the coordinate origin, i.e. $O(0,0,0)$. Each item is represented by its position of back-left-bottom corner coordinate $(x_i, y_i, z_i)$ and dimensions of depth $d_i$, width $w_i$ and true height $h_i^*$ as shown in Fig. 4.

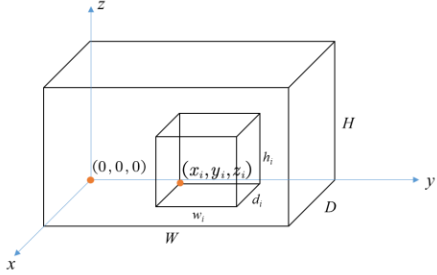

**Fig. 4.** The representations of an item and a bin in the three-dimensional coordinate system.

Our objective is to maximize the initial volume of all the items $IV$ packed into each bin, which can be expressed by the following formulas:

$$\max\ IV = \sum_i d_i w_i h_i x_i, \quad \forall i = \{1, 2, ..., n\} \tag{3}$$

where $x_i$ is the decision variable. When item $i$ is to be packed in to the bin, $x_i = 1$, otherwise, $x_i = 0$.

A feasible packing solution in this paper must satisfy the following constraints:

1) *Weight limit*: The total weight of loaded items in the bin cannot exceed the maximum weight limit of the bin.
2) *Space limit*: Items can only be placed inside but not outside the container, that is, the three dimensions width, depth and height of the loaded item cannot exceed those of the bin.
3) *Orthogonality*: Each item must be loaded orthogonally into the bin, that is, the faces of the loaded item are parallel to the walls of the bin.
4) *No overlapping*: Items should not be loaded with overlapping.
5) *Stability*: The loaded items cannot be suspended in the air. Their bottom faces must touch either the top faces of other items or the floor of the bin underneath.
6) *Orientation*: Each item has 6 rectangular facets, but there are only 3 distinct facets because "opposite" facets are identical. Each of the three facets can be rotated orthogonally to obtain a new dimension of the item. Thus, each item can have 6 different rotation types, which can be obtained by rotating about the $x$, $y$ and/or $z$ axis as shown in Fig. 5. and Table 1.

**Table 1.** 6 different rotation types obtained by rotating about different axis.

| Rotation Type | First axis to rotate about | Second axis to rotate about |
|---|---|---|
| 0 | - | - |
| 1 | Z | - |
| 2 | X | Y |
| 3 | Y | - |
| 4 | X | Z |
| 5 | X | - |

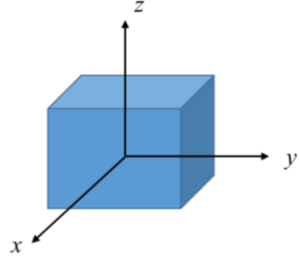

**Fig. 5.** Objects can be rotated about the three axes of symmetry.

## 4. PROPOSED ALGORITHM

### 4.1. THE PROPOSED DYNAMIC-VOLUME-BASED PACKING ALGORITHM

In this paper, we propose a new method for implementing a bottom-left-fill packing strategy, which considers height change due to compression during the packing progress of items and prevents items from hanging in the air. Only one bin will be opened at a time and the algorithms use a series of *pivot points* where to pack the item. Pivot point is an $(x, y, z)$ coordinate which represents a point in a particular three-dimensional bin where an attempt to pack an item will be made. The back-left-bottom corner of the item will be placed at the pivot point. The first and the only one pivot point in an empty bin is always $O(0, 0, 0)$. When item $i$ is placed with its back-left-bottom corner in the pivot point $(x, y, z)$, for example, $O(0, 0, 0)$, it will generate a series of potential pivot points for the items that have not been packed, which includes $pX(x + d_i, y, z)$, $pY(x, y + w_i, z)$ and $pZ(x, y, z + h_i)$ as shown in Fig. 6.

To prevent items from hanging in the air after being packed, each pivot point will be considered whether it is feasible. A pivot point is invalid if it's dangling, so we need to select a new valid pivot point. The new valid pivot point is generated by looking from top to bottom and take the mapping point on the top surface of the first item below pivot point $pX$ ($pY$), as shown by the orange dots in Fig. 7. What should be particular noted is that the pivot point to the left of new valid pivot point, as shown by the blue dots in Fig. 7., will still be given priority. In this case, the item will not be able to be hanging in the air. All the new pivot points will be appended to the list of pivots. The pivot points in list of pivots are sorted in the bottom-left-deepest order, that is, the pivot points are scanned according to lowest $z$ values, breaking ties by lowest $y$ and then by lowest $x$ values.

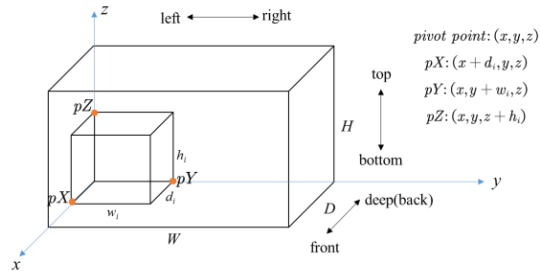

**Fig. 6.** Bottom-left-fill packing strategy: when item $i$ is placed with its back-left-bottom corner in the pivot point $(x, y, z)$, it will generate a series of potential pivot points for the items that have not been packed.

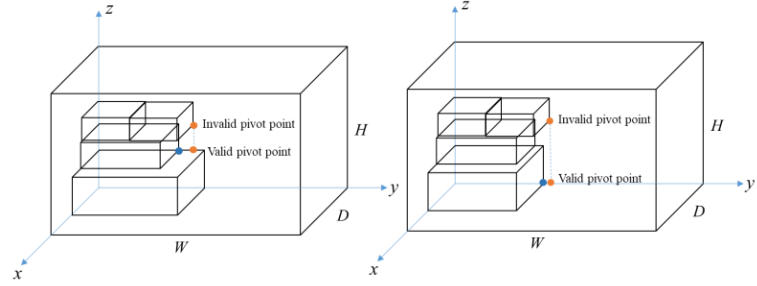

**Fig. 7.** The orange dots represent the mapping of invalid points to valid points. The blue dots are given priority to be considered.

For each item, the facet with the largest area of the item is defined as the bottom facet, which helps to ensure the stability of the item to a certain extent. The items to be packed into the bin will be first sorted according to the bottom facet area from large to small, and for items with the same bottom facet area they will be sorted according to their compressibility. This sorting approach is more in line with our actual experience, that is to give priority to items with larger bottom facet area and compressibility packed at the bottom of the bin, which will make it easier to load more items. When packing each item, we prioritize the bottom-down rotation type and then traverse the list of pivots until the item can be packed at a pivot point, otherwise item is rotated and re-traverse the list of pivots until we have tried all 6 possible rotation types. If the item still cannot be packed at the pivot point after being rotated, then we move on to packing the next item and add the unpacked item to a list of items that cannot be packed in to the bin. Algorithm 1 gives the pseudo-code of the proposed dynamic-volume-based loading algorithm, where the list of pivots stores a series of tuples each of which used to store the pivot point and the item under the corresponding point, for example, ((0,0,0), None) for the origin.

---

**Algorithm 1 Dynamic-volume-based packing algorithm**

1: Initialize pivot point $p \leftarrow [0,0,0]$, the item under current pivot point $a \leftarrow \varnothing$, the Pivot_list $P \leftarrow \{(p,a)\}$

2: Three new pivot points $pX, pY, pZ$ generated after packing an item in pivot point $p$

3: Sort the Items $I$ according to maximum floor area $S$ and compressibility $c$: sort($I$, key= $S, c$)

4: for each item $i \in I$ do

5:    for each rotation_type do

6:        for each $(p,a) \in P$ do

7:            if put_item($i, p$, rotation) == True: then

8:                $P \leftarrow P/\{(p,a)\}$

9:            if $pX$ or $pY$ goes beyond the surface of the item below then

10:               $height, a \leftarrow (height, a)$ of the upper surface of the first object found straight down from $pX$ or $pY$

12:               $P \leftarrow P \cup \{([pX(pY)[0], pX(pY)[1], height], a)\}$

13:           else

14:               $P \leftarrow P/\{(pX, a)\}$

15:               $P \leftarrow P/\{(pY, a)\}$

16:               $P \leftarrow P/\{(pZ, a)\}$

17:               sort($P$, key $=p[2], p[1], p[0]$)

18:           end if

19:        end if

20:        end for

21:    end for

22: end for

**Algorithm. 1.** The proposed dynamic-volume-based loading algorithm.

### 4.2. MODEL OF 3D-BPP OF DEFORMABLE ITEMS WITH DYNAMIC VOLUME CHANGE

We establish the 3D-BPP model of irregular deformable items with dynamic volume change that simulates the compression process of items and estimate whether the item can be packed into the bin. The model provides a general framework for handling the 3D-BPP of deformable items and can be used to evaluate the performance of different heuristic algorithms.

For item $i$ with a certain rotation type and a pivot point, the pivot point and the sizes of item are regarded as the position and the dimension of the item respectively. When trying to put current item $i$ in the top of some other items in the bin, it's necessary to calculate whether the total true height will exceed that of bin. The position, the total weights of items placed on the current item $i$, true compressibility and true height of each item under item $i$ considering should be updated after being compressed. If item $i$ cannot be packed into the bin even after the process of compression, it means that it will totally not fit in this bin. Algorithm 2 gives the pseudo-code of the model of how to estimate whether current item can fit into the bin.

---

**Algorithm 2 Estimate if the item can fit into the bin**

1: Function: Put_item($i$, $(p,a)$, rotation_type)
2: Initialize the item under $i$ is $n_i \leftarrow a$, the item on $i$ is $m_i \leftarrow \varnothing$ the total weight of all the items on current item $c$ is $u_c \leftarrow 0$
3: if $i$ is wide or deep beyond the edge of the bin then
4:     return False
5: end if
6: if there is no item under $p$ in the bin then
5:     if ($i$ is high beyond the top of the bin) or (intersect with items in the bin) then
6:         return False
7:     end if
8: end if
9: while $n_i\ != None$ do
10:     $n_i \leftarrow n_{n_i}$
12:     $u_{n_i} \leftarrow u_{n_i} + w_{n_i}$
13: end while
14: item_list $L \leftarrow [n_i]$
15: while $L$ is not *NULL* do
16:     current item $c \leftarrow L[0]$
17:     calculate and update the actual position, compressibility and true height of current item
18:     $L \leftarrow L \cup \{m_c\}$
19:     $L \leftarrow L/\{c\}$
20: end while
21: if (the new height beyond the top of the bin) or (intersect with items in the bin) then
22:     return False
23: else
24:     return True
25: end if

**Algorithm. 2.** The model of how to estimate whether current item can fit into the bin.

## 5. EXPERIMENTS

In this section, we summarize the experimental results. The proposed algorithm and model are implemented in Python3. All the computational experiments are carried out on a laptop with 2.80GHz CPU and RAM 16 GB.

### 5.1. EXPERIMENTAL SETTINGS

Due to the lack of shared test data for these kinds of studies, we test our algorithm and model using data generated randomly, which includes different sizes of bins (as Table 2 shows) and 4 types of items (as Table 3 shows). Items are divided into four categories: green vegetables(Type 0), rice(Type 1), melons and fruits(Type 2), and other categories(Type 3), corresponding to different compressibility and maximum compression ratio. Table 4 shows the examples of items used in experiment, where only one single bin will be considered at each time.

### 5.2. EXPERIMENTAL RESULTS

To verify the performance of the new approach, our method is tested on the data set comparing the initial volume and number of all the packed items and space utilization for each bin with and without compression. As shown in Fig. 8.(a), the space utilization with compression under bins of different sizes is higher than that without compression. Fig. 8.(b) demonstrates that the bin can hold more items when taking compression into consideration. Table 5 reports the results of the initial volume and true volume (after being compressed) of total items in the bin, the number of packed items and space utilization for different sizes of bins by the proposed method, which shows good behaviour of improving initial volume of total items.

**Table 2.** Four different types of bins

| Bin Type | Depth(cm) | Width(cm) | Height(cm) | Max weight(kg) | Quantity |
|---|---|---|---|---|---|
| Small bin | 40 | 40 | 35 | 55 | 1 |
| Medium bin | 50 | 45 | 40 | 65 | 1 |
| Large bin | 60 | 50 | 45 | 80 | 1 |
| Larger bin | 70 | 65 | 60 | 100 | 1 |

**Table 3.** Four different types of items

| Item Type | Compressibility | Maximum compression ratio |
|---|---|---|
| 0 (Green vegetables) | 0.1 | 0.3 |
| 1 (Rice) | 0.03 | 0.2 |
| 2 (Melons and fruits) | 0.01 | 0.1 |
| 3 (Other categories) | 0 | 0 |

**Table 4.** All the candidate items

| Item Name | Depth(cm) | Width(cm) | Height(cm) | Weight(kg) | Quantity | Type |
|---|---|---|---|---|---|---|
| Chinese Cabbage | 25 | 12 | 12 | 1.2 | 50 | 0 |
| Little Cabbage | 18 | 8 | 8 | 0.8 | 50 | 0 |
| Rice | 45 | 40 | 8 | 5 | 5 | 1 |
| Millet | 35 | 30 | 8 | 2.5 | 5 | 1 |
| Bebe Pumpkin | 10 | 10 | 7 | 0.3 | 50 | 2 |
| Potato | 12 | 5 | 5 | 0.1 | 50 | 2 |
| Eggs | 30 | 20 | 20 | 1.6 | 5 | 3 |

### 5.3. VISUALIZATION BIN PACKING

In order to show the results more intuitively, we use python for visualization. Fig. 9. visualizes the effect of different sizes of bins and dynamic loading animation is uploaded on the website https://github.com/3D-irregular-bin-packing/Video/tree/main. It can be seen that our visualization can conveniently give the staff suitable solutions in real time in practice.

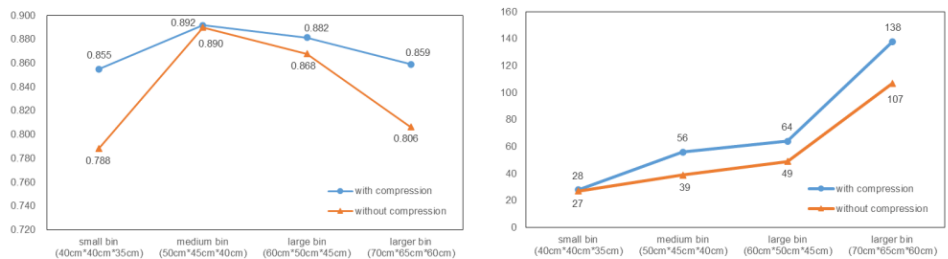

(a) Comparison of space utilization       (b) Comparison of the number of items packed in the bin

**Fig. 8.** Comparison of the results with and without compression under different sizes of bins

## 6. CONCLUSION

We investigate the 3D BPP for irregular-shaped items. In the context of fresh food delivery in this work, instead of assuming the items are rigid and no deformation will occur during the packing procedure, we consider that the volume of the items changes dynamically while packing due to extrusion. We propose a dynamic-volume-based heuristic to solve the studied problem. Computational results reveal that the space utilization is improved after incorporating the volume change issues.

Future work may focus on proposing the mixed integer programming model to obtain the optimal solution and serve as benchmarks to estimate the solution quality of heuristic solutions yielded in this work.

**Table 5.** Packing situation of each bin

| Bin Type | Initial volume of total items | True volume of total | The number of items packed | Space |
| --- | --- | --- | --- | --- |
| Small bin | 51412.00 | 47881.35 | 28 | 0.855 |
| Medium bin | 85200.00 | 80259.48 | 56 | 0.892 |
| Large bin | 155132.00 | 143729.04 | 64 | 0.882 |
| Larger bin | 264376.00 | 234505.64 | 138 | 0.859 |

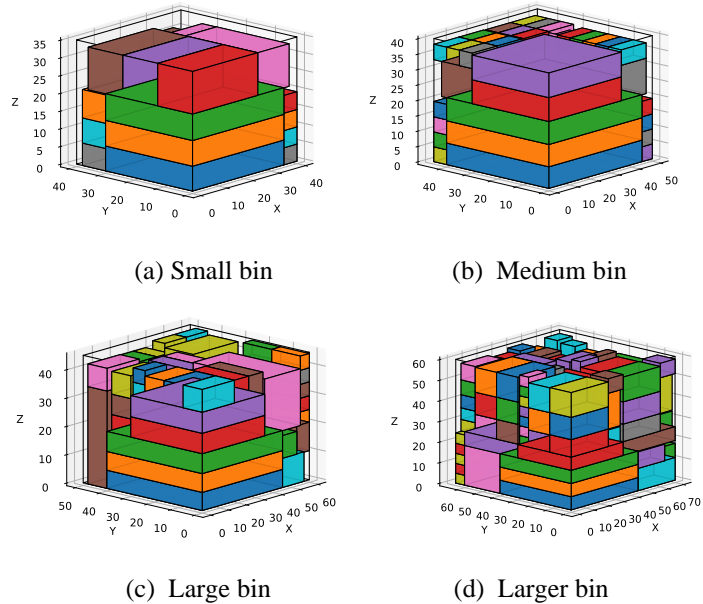

(a) Small bin    (b) Medium bin

(c) Large bin    (d) Larger bin

**Fig. 9.** Visualization of items with compression

**ACKNOWLEDGEMENTS**

This research was partially supported by the Shenzhen Science and Technology Innovation Commission (JCYJ20210324135011030, WDZC20200818121348001), Guangdong Pearl River Plan (2019QN01X890), and National Natural Science Foundation of China (Grant No. 71971127), and Shenzhen Yue Hou Technology Co., Ltd.

**REFERENCES**

1. Kaabi J, Y Harrath, Bououdina H E, et al. Toward Smart Logistics: A New Algorithm for a Multi-Objective 3D Bin Packing Problem[C]// Smart Cities Symposium 2018. 2018.
2. Martello, Silvano, Pisinger, et al. THE THREE-DIMENSIONAL BIN PACKING PROBLEM.[J]. Operations Research, 2000.

3. Lodi A , Martello S , Vigo D . Heuristic algorithms for the three-dimensional bin packing problem[J]. European Journal of Operational Research, 2002, 141(2):410-420.
4. Gzara F , Elhedhli S , Yildiz B C . The Pallet Loading Problem: Three-dimensional Bin Packing with Practical Constraints[J]. European Journal of Operational Research, 2020.
5. Jiang J , Cao L . A hybrid simulated annealing algorithm for three-dimensional multi-bin packing problems. IEEE, 2012.
6. Elhedhli S , Gzara F , Yildiz B . Three-Dimensional Bin Packing and Mixed-Case Palletization[J]. INFORMS Journal on Optimization, 2019, 1(4):ijoo.2019.0013.
7. Martello S , Pisinger D , Vigo D , et al. Algorithm 864: General and robot-packable variants of the three-dimensional bin packing problem[J]. Acm Transactions on Mathematical Software, 2007, 33(1):7.
8. Fekete S P , Schepers J , Veen J . An exact algorithm for higher-dimensional orthogonal packing[J]. 2006.
9. Hifi M , Kacem I , S Nègre, et al. A Linear Programming Approach for the Three-Dimensional Bin-Packing Problem[J]. Electronic Notes in Discrete Mathematics, 2010, 36(none):993-1000.
10. Junqueira L , Morabito R , Yamashita D S . Three-dimensional container loading models with cargo stability and load bearing constraints[J]. Computers & Operations Research, 2012, 39( 1):74-85.
11. Schyns, M, Limbourg, et al. A mixed integer programming formulation for the three-dimensional bin packing problem deriving from an air cargo application[J]. International transactions in operational research: A journal of The International Federation of Operational Research Societies, 2016, 23(1/2):187-213.
12. George J A , Robinson D F . A heuristic for packing boxes into a container[J]. Computers & Operations Research, 1980, 7(3):147-156.
13. Scheithauer, Guntram. (1991). A Three-dimensional Bin Packing Algorithm.. Elektronische Informationsverarbeitung und Kybernetik. 27. 263-271.
14. Bischoff E E , Ratcliff M . Issues in the development of approaches to container loading[J]. Omega, 1995, 23(4):377-390.
15. Eley M . Solving container loading problems by block arrangement[J]. European Journal of Operational Research, 2002, 141(2):393-409.
16. A T G C , B G P , B R T . TS 2 PACK : A two-level tabu search for the three-dimensional bin packing problem[J]. European Journal of Operational Research, 2009, 195( 3):744-760.
17. Jiang J , Cao L . A hybrid simulated annealing algorithm for three-dimensional multi-bin packing problems. IEEE, 2012.
18. Jiang, Y., Cao, Z., & Zhang, J. (2021). Solving 3D Bin Packing Problem via Multimodal Deep Reinforcement Learning. AAMAS.
19. Martinez-Sykora A , Alvarez-Valdes R , Bennell J , et al. Matheuristics for the irregular bin packing problem with free rotations[J]. European Journal of Operational Research, 2017:S0377221716307950.
20. Abeysooriya R P , Bennell J A , Martinez-Sykora A . Jostle heuristics for the 2D-irregular shapes bin packing problems with free rotation[J]. International Journal of Production Economics, 2017, 195(jan.):12-26.
21. Agrawal P K . Minimising trim loss in cutting rectangular blanks of a single size from a rectangular sheet using orthogonal guillotine cuts[J]. European Journal of Operational Research, 1993, 64(3):410-422.
22. Kim J Y , Kim Y D . Graph theoretic heuristics for unequal-sized facility layout problems[J]. Omega, 1995, 23(4):391-401.